\documentclass[%
   aps,
   prb,
   reprint,
   amsmath,%
   amssymb,%
   superscriptaddress,%
   showpacs,%
   showkeys,%
   floatfix,%
]{revtex4-1}

\usepackage{graphicx}

\newcommand{\tc}{\ensuremath{T_\text{c}}}
\newcommand{\xray}{X-ray}
\newcommand{\bragg}[3]{\ensuremath{#1\,#2\,#3}}
\newcommand{\etal}{\textit{et~al}.}
\newcommand{\vektor}[1]{\ensuremath{\mathbf{#1}}}
\newcommand{\sg}[4]{\ensuremath{#1\,#2\,#3\,#4}}
\newcommand{\disangle}{\ensuremath{(\gamma^\ast-90^\circ)}}

\begin{document}
\title{Tetragonal to Orthorhombic Transition of GdFeAsO Studied by
Single-Crystal Synchrotron X-Ray Diffraction}

\author{F.~Nitsche}
\affiliation{Department of Chemistry and Food Chemistry, Technische
Universit\"{a}t Dresden, D-01062 Dresden, Germany}
\author{Th.~Doert}
\email{thomas.doert@chemie.tu-dresden.de}
\affiliation{Department of Chemistry and Food Chemistry, Technische
Universit\"{a}t Dresden, D-01062 Dresden, Germany}
\author{M.~Ruck}
\affiliation{Department of Chemistry and Food Chemistry, Technische
Universit\"{a}t Dresden, D-01062 Dresden, Germany}
\affiliation{Max Planck Institute for Chemical Physics of Solids, D-01187
Dresden, Germany} 

\date{\today}

\begin{abstract}
A study of the tetragonal to orthorhombic phase transition of GdFeAsO is
presented. Planes of the reciprocal space were reconstructed form 
single-crystal synchrotron X-ray diffraction data. By cooling below the
structural transition temperature splitting of the Bragg reflections was
observed corresponding to four different twin domain orientations. A 
model was developed to quantify the distortion of the lattice 
from the position of the splitted reflections relative to each other.
Constrained 2D-Cauchy fits of several splitted reflections provided positions of
the reflections. The influence of the structural distortion was detectable
already above the structural transition temperature hinting at fluctuations in
the tetragonal phase.
\end{abstract}

\pacs{%
61.50.Ks  
61.66.Fn, 
74.62.Bf, 
74.70.Xa, 
}

\keywords{Fe-based superconductors, synchrotron x-ray diffraction, phase
transition, twinning, single crystal}

\maketitle

\section{Introduction}
With the discovery of high-temperature superconductivity in layered iron
containing compounds (iron pnictides) \cite{Kamihara2006, Kamihara2008,
Paglione2010}, a new system of phases was added to the study of
unconventional superconductivity. Since the understanding of the phenomenon
of high-\tc\ superconductivity is still limited, the question of the
differences and similarities towards other high-\tc\ compounds such as the
cuprates arises\cite{Tohyama2012}.

The cuprates as well as many iron pnictides become superconducting by doping of
a non-superconducting antiferromagnetic ``parent'' compound. Whereas
the superconductivity in the cuprates is believed to originate from the
Mott-insulator properties of the ``parent'' compound, the iron
pnictides do not exhibit this electronic correlation.
However, a common feature is the antiferromagnetic ordering, which has to be
suppressed to achieve superconductivity and is thus considered significant. 

Iron
pnictides, such as $RE$FeAsO, ($RE$ = rare-earth metal) and $A$Fe$_2$As$_2$ 
($A$ = Ca, Sr, Ba, and Eu), show antiferromagnetic transitions accompanied by 
structural distortions from tetragonal to orthorhombic symmetry upon cooling.
While the structrual transition and the magnetic ordering occur simultaneously in
$A$Fe$_2$As$_2$\cite{Rotter2008, Tegel2008}, 
the structural transition
in $RE$FeAsO compounds 
is observed at slightly higher temperatures than the magnetic 
ordering\cite{delaCruz2008, Klauss2008, Luo2009, Jesche2010,
Klingeler2010}. 

\begin{figure}
\centering
\includegraphics{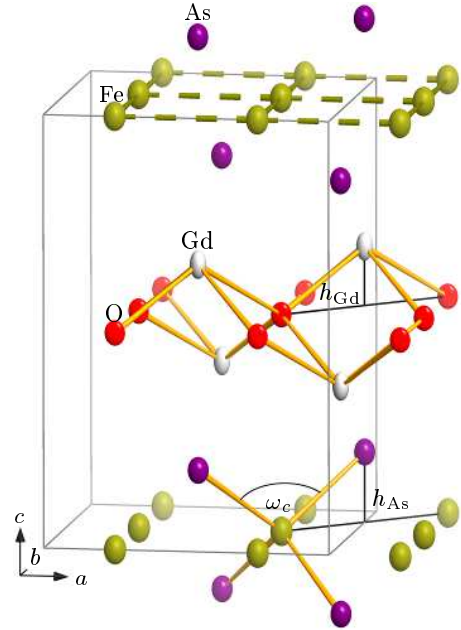}
\caption{(Color online) Low-temperature crystal structure of 
GdFeAsO at $50\,$K (space group \sg{C}{m}{m}{e}, No.~$67$). The displacement
ellipsoids represent $99.9\,$\% probability. Solid lines between
iron atoms indicate shortest Fe-Fe distance, dashed lines are slightly longer.
$h_\text{Gd}$ is the height of gadolinium ions above the oxygen layer,
$h_\text{As}$ is the height of the arsenic ions above the iron layer, 
and $\omega_c$ is the As-Fe-As angle with the angle bisector parallel to
the $c$-direction.  \label{fig:unitcell}}
\end{figure}

The $RE$FeAsO room-temperature phases of ZrCuSiAs-type structure 
(space-group type \sg{P}{4/n}{m}{m}) 
undergo a
\emph{translationengleiche} transition of index 2 
to the low-temperature structure
with space-group type \sg{C}{m}{m}{e}\cite{delaCruz2008}
(Fig.~\ref{fig:unitcell}, for structural
relations see also Ref.~\onlinecite{Johrendt2011}). 
A similar 
transition occurs in the $A$Fe$_2$As$_2$ compounds (\sg{I}{4/m}{m}{m} to
\sg{F}{m}{m}{m}). 
Tanatar \etal\cite{Tanatar2009} and Blomberg
\etal\cite{Blomberg2012} studied the formation of lamellar transformation twins
by single-crystal \xray\ diffraction of $A$Fe$_2$As$_2$ and the influence of the
domain structure on the
anisotropy of the resistivity in directions of the basal plane.
Due to the analogous mechanism of the structural transition for $RE$FeAsO, the
question arises, if the twin formation is similar to $A$Fe$_2$As$_2$.

Here, we present the first single-crystal \xray\ diffraction study of the
structural evolution and the twin formation of $RE$FeAsO compounds with
GdFeAsO as representative. 
Doped GdFeAsO has one of the highest \tc\ ($56\,$K) in this class of
superconductors\cite{Wang2008}, which has been associated with the vanishing
distortion of the FeAs$_4$ tetrahedra\cite{Lee2008}.
This concept has been the starting point of an ongoing investigation on
the structural changes associated with the transition to the superconducting
phase and its physical implications. 

\section{Experimental}
Preparation and characterization of the crystal used for
single-crystal synchrotron \xray\ diffraction has been described
elsewhere\cite{Nitsche2010}. Diffraction data for this study were recorded with
a $165\,$mm MAR-CCD detector mounted on a Huber four-circle diffractometer at
beamline D3 at DESY. The measurements were performed in the
temperature range of $50$--$300\,$K using an open flow Oxford Diffraction
Helijet cryostat.
$\varphi$-scans were recorded with an increment of~$1^\circ$, a detector
distance of $54.4\,$mm, and a wavelength of $0.49741\,$\AA. The detector
distance and the wavelength were refined
by comparing the experimental lattice parameters of a standard
corundum single-crystal to literature data.
The raw frames from the MAR-CCD detector were converted using the APEXII 
suite\cite{APEX2}. Integration and corrections for oblique
incidence and polarization were performed within SAINT+\cite{SAINT}. For data
reduction and absorption correction SADABS\cite{Sheldrick2008} was used.
The
structures were solved and refined with SHELXS and 
SHELXL\cite{Sheldrick2007}. Details concerning the structure analysis of
GdFeAsO at different temperatures can be obtained from the
Fachinformationszentrum Karlsruhe\cite{Note1}.

\section{Results and Discussion}
\begin{figure}
\centering
\includegraphics{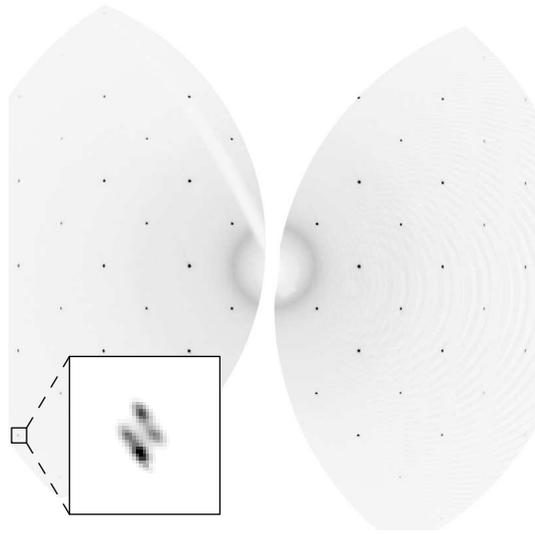}
\caption{Reconstructed \bragg{h}{k}{0} plane in monoclinic indexing of
GdFeAsO at $50\,$K. \emph{Inset}: Magnification of the
\bragg{\bar{6}}{\bar{4}}{0}
reflections.\label{fig:splitting}} 
\end{figure}

When cooling below the temperature of the structural transition,
$T_\text{s}\approx 115\,$K, 
splitting of reflections at high diffraction angles was observed in the
synchrotron experiment. 
To assess the metric
distortion during the structural transition, a program 
was written in MATLAB\cite{MATLAB2011}, that allows reconstruction of 
planes of the reciprocal space with high lateral resolution
(Fig.~\ref{fig:splitting}).  

\begin{figure*}
\centering
\includegraphics{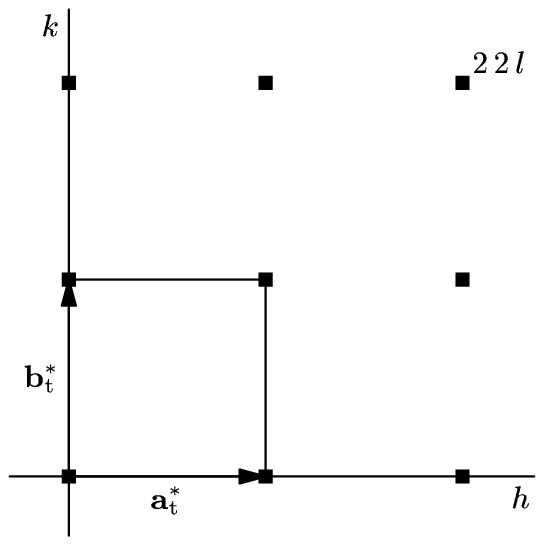}
\hspace{0.5cm}
\includegraphics{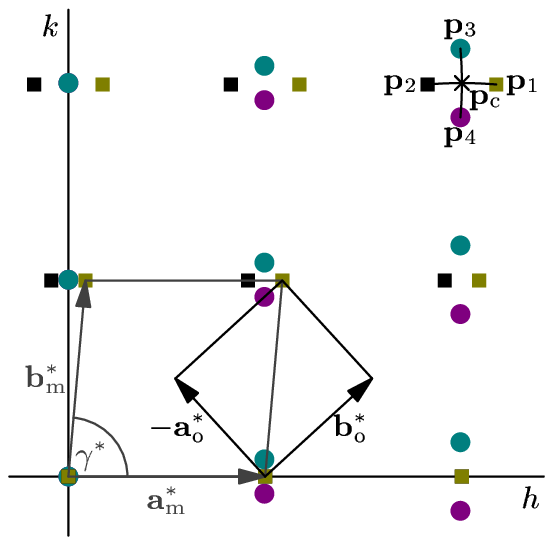}\\\vspace{0.5cm}
\hspace{0.1cm}\raisebox{0.5cm}{\includegraphics{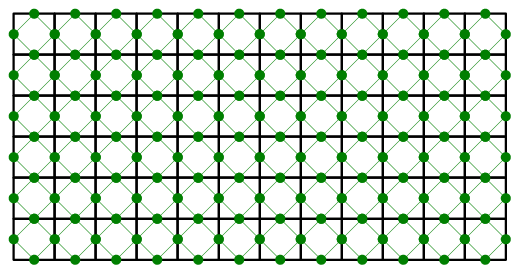}}
\hspace{0.8cm}
\includegraphics{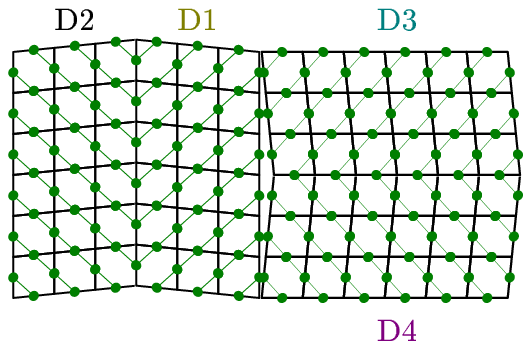}
\caption{(Color online) Scheme of the tetragonal (left) to orthorhombic (right) distortion in
analogy to Ref.~\onlinecite{Tanatar2009}.
\emph{Upper panels}: Schematic reflection patterns for $a^\ast\,b^\ast$-planes
with $l \neq 0$. \emph{Lower panels}: Real space
domains. Black lines represent the primitive unit cells, green points are
the positions of iron atoms and green lines show shortest Fe-Fe
distance. D1 to D4 denote the domain orientations corresponding to p1
to p4 in the upper right panel.\label{fig:domains}}
\end{figure*}

The metric distortion of the structure is the only degree of freedom generated
by the structural transition. A model has been developed, which allows to 
determine the metrical distortion from the observed splitting of reflections.
For convinient comparison of the tetragonal (indicated by subscript t) and 
orthorhombic (subscipt o) structures, the
$C$-centered orthorhombic cell is transformed into a monoclinic (subscript m)
primitive cell (upper right panel in Fig.~\ref{fig:domains}) by 
\begin{eqnarray*}
\vektor{a_\text{m}} = \frac{1}{2}\left(\vektor{a_\text{o}} +
\vektor{b_\text{o}}\right)
\text{ and }
\vektor{b_\text{m}} = \frac{1}{2}\left(- \vektor{a_\text{o}} +
\vektor{b_\text{o}}\right) \\
\text{with }
\left|\vektor{a_\text{m}}\right| = \left|\vektor{b_\text{m}}\right|
\text{ and }
\left|\vektor{a_\text{o}}\right| > \left|\vektor{b_\text{o}}\right|
\text{.}
\end{eqnarray*}
The distortion of the structure is then defined by the angle 
$\gamma > 90^\circ$ (i.e. $\gamma^\ast < 90^\circ$ in reciprocal space).

The twin domains join in common planes that are indexed $(1\,1\,0)$ and
$(1\,\bar{1}\,0)$ in the C-centered cell, i.e. $(1\,0\,0)$ and $(0\,1\,0)$ in
the primitive cell. Thus four different orientations of the domains in the
low-temperature phase result. In the diffraction pattern fourfold splitting of
reflections in general positions can be observed, especially at higher
diffraction angles (Inset in Fig.~\ref{fig:splitting}).
The position vectors $\vektor{p}_1$, $\vektor{p}_2$,
$\vektor{p}_3$, and $\vektor{p}_4$ of the reflections belonging to the four
domains can be derived from their common center $\vektor{p}_c$ 
and the distortion angle $\gamma^\ast$ (Fig.~\ref{fig:domains}) by 
\begin{eqnarray*}
\vektor{p}_1 = \begin{pmatrix}1&-\sin\disangle\\0&\cos\disangle\end{pmatrix} \, \vektor{p}_c
\text{, }&&
\vektor{p}_2 = \begin{pmatrix}1&\sin\disangle\\0&\cos\disangle\end{pmatrix} \, \vektor{p}_c
\text{, }\\
\vektor{p}_3 = \begin{pmatrix}\cos\disangle&0\\-\sin\disangle&1\end{pmatrix} \, \vektor{p}_c
\text{, }&&
\vektor{p}_4 = \begin{pmatrix}\cos\disangle&0\\\sin\disangle&1\end{pmatrix} \, \vektor{p}_c
\text{. }
\end{eqnarray*}

Blomberg \etal\cite{Blomberg2012} detwinned BaFe$_2$As$_2$ crystals by tensile
stress and showed that the relative positions of the reflections is
dependent on the domain fractions. If, for example, the domain fraction of D4 in
Fig.~\ref{fig:domains} increases, the orientation of the two sets of domains
changes towards each other.
In the geometrical model described here, the correlation of the reflection
positions to their intensity is achieved by linear scaling of the transformations
shown above. This approximation is acceptable due to the small differences of
the intensities and the local fit of the reflections.
In summary, the positions of the reflections are
constrained to their common center $p_c$ of all the four reflections, the monoclinic
distortion angle $\gamma^\ast$, and the intensities of the reflections.
This enables the parameter refinement of overlapping reflections.

\begin{figure}
\centering
\includegraphics{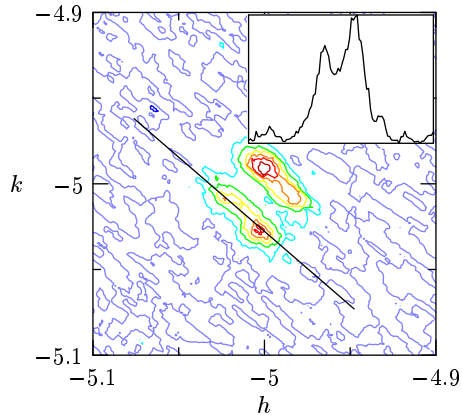}
\caption{(Color online) Contour plot of the $\bar{5}\,\bar{5}\,0$ reflection in
\bragg{h}{k}{0} plane. \emph{Inset}: 1D-section of the 2D-distribution
as indicated by the black line in the contour plot.\label{fig:slice}}
\end{figure}

To quantify the distortion, an algorithm was written
in MATLAB to fit the splitted reflections
by a linear combination of four elliptical
2D-Lorentz (Cauchy) functions 
(concerning multivariate
t-distributions see Ref.~\onlinecite{Kotz2004}, 2D-Gaussian fit of neutron data see
Refs.~\onlinecite{McIntyre1986, McIntyre1988}). A Lorentzian type profile function was
chosen empirically by investigating 1D-sections of the 2D-distribution (see
Fig.~\ref{fig:slice}).
The fitting parameters were the center of the splitted reflections
$\vektor{p}_\text{c}$, the distortion angle ($\gamma^\ast$), the
ellipticity of the reflections, the
width of the 2D-Lorentz distributions, and the individual intensities of the
reflections. 

For the structural investigation $20$ reflections of each measurement,
exhibiting reflection splitting, were fitted to obtain an average distortion
angle $\gamma^\ast$ and a corresponding standard deviation. The reflection data
were
integrated with a tetragonal cell since splitting was not wide enough to
perform multidomain integration. Using the tetragonal lattice parameters, which
correspond to $\vektor{p_\text{c}}$, and the obtained distortion angle, the
orthorhombic lattice parameters were obtained by
\[
a_\text{o} = 2\,a_\text{t}\,\sin\left(\frac{\gamma}{2}\right)
\text{ and }
b_\text{o} = 2\,a_\text{t}\,\cos\left(\frac{\gamma}{2}\right)\text{.}
\]

All reflections could be fitted with this model. No signs for a deviation from 
the low-temperature orthorhombic structure towards true monoclinic symmetry was
observed. 
Furthermore, no superstructure reflections could be identified in the
reconstructed images of the reciprocal space. This confirms the
tetragonal to orthorhombic transition established by powder
diffraction\cite{delaCruz2008}.

\begin{figure}
\centering
\includegraphics{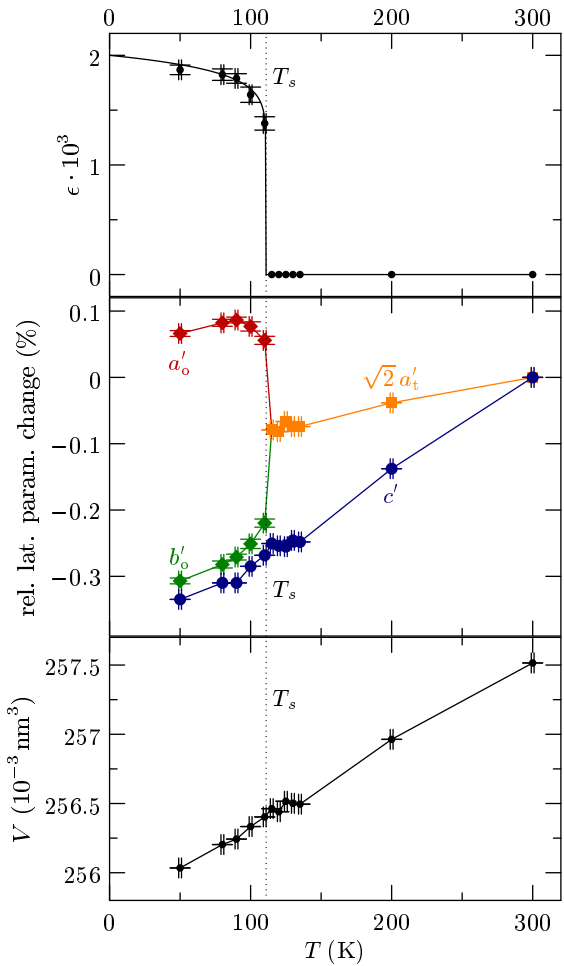}
\caption{(Color online) Temperature dependence of selected structural parameters of
GdFeAsO single crystal. Error bars represent $\pm\sigma$. \emph{Upper panel}: Spontaneous
deformation fitted by a power law. \emph{Middle panel}: Deviation of the lattice
parameters relative to $300\,$K. 
(%
$a^\prime_\text{t} = a_\text{t}/a_\text{t}(300\,\text{K})-1$, 
$a^\prime_\text{o} = a_\text{o}/\left[\sqrt{2}\,a_\text{t}(300\,\text{K})\right]-1$, 
$b^\prime_\text{o} = b_\text{o}/\left[\sqrt{2}\,a_\text{t}(300\,\text{K})\right]-1$, and
$c^\prime = c/c(300\,\text{K})-1$%
).
\emph{Lower Panel}: Cell volume of the orthorhombic phase and
doubled volume of the tetragonal phase for comparison. Lines are
guides for the eye in the middle and lower panel.\label{fig:plots}}
\end{figure}

The order parameter of the tetragonal to orthorhombic transition is
proportional to the spontaneous deformation $\epsilon$ and can be fitted by a
power law (upper panel in Fig.~\ref{fig:plots}), 
\[
\epsilon = \frac{a-b}{a+b} = \tan\left(\frac{\gamma-90^\circ}{2}\right) 
= A\,\left(1-\frac{T}{T_\text{s}}\right)^\beta 
\]
with the critical exponent $\beta = 0.072(1)$ and a transition temperature
of $T_\text{s} = 111(1)\,$K.

The critical exponent is very small but comparable to the order parameters
found for SrFe$_2$As$_2$ and EuFe$_2$As$_2$ ($0.098(1)$ and
$0.112(1)$, respectively\cite{Tegel2008}), where structural and magnetic 
transition occur
at the same temperature. Tegel \etal\ explained the small critical exponent by
an ordering following the two-dimensional Ising
model of the iron subsystem, which should yield a critical exponent of $1/8$. In the
case of GdFeAsO the structural transition occurs at temperatures above the
magnetic transition,
which could be the reason for the even lower critical exponent determined,
indicating that the structural transition of GdFeAsO is close to first order
(concerning multiferroic transformation and the connection to first order
phase transition see also
Refs.~\onlinecite{Jacobs2000, Salje2011}).

The structural transition temperature $T_\text{s} = 111(1)\,$K obtained from
the fit is lower than determined by Luo \etal\cite{Luo2009} with powder \xray\
diffraction and resistivity measurements. This deviation might be caused by a
systematic temperature error from the open flow helium cryostat used for this
single-crystal \xray\ diffraction study. 

The lattice parameters and the volume of the tetragonal phase (middle and lower
panel of Fig.~\ref{fig:plots}) decrease in the course of cooling. However,
there are hints that there is an anomaly just above the structural transition
(e.g. different slope for $c$ below $140\,$K),
as it was also observed for the thermal expansion of GdFeAsO by Klingeler
\etal\cite{Klingeler2010} and assigned to a competition of differently ordered
phases.

\begin{figure}
\centering
\includegraphics{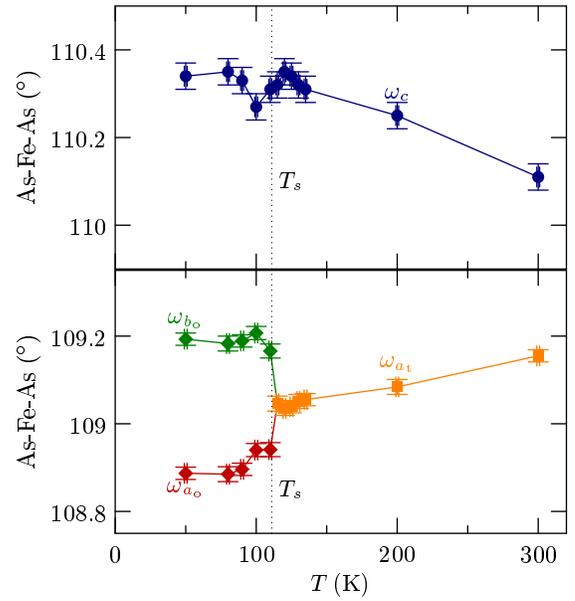}
\caption{(Color online) Temperature dependence of the As-Fe-As angles $\omega$ with its
bisector parallel to the direction in subscript. Note the relation
$\cos\omega_{a_\text{t}} = -\frac{1}{2}\,\left(1+\cos\omega_c\right)$.
Lines are guides
for the eye, error bars represent $\pm\sigma$.\label{fig:angles}}
\end{figure}

Interestingly, the height of the gadolinium atom above the oxygen layer
($h_\text{Gd}$ in Fig.~\ref{fig:unitcell}) remains
almost constant over the whole temperature range, while the arsenic atom height
over the iron layer ($h_\text{As}$) diminishes with decreasing lattice
parameter $c$.  The angle $\omega_\text{c}$ of the coordination polyhedron
around Fe (Fig.~\ref{fig:angles})
with its bisector parallel to the $c$-direction first increases upon cooling and
then becomes almost constant already above the structural
transition and does not change much upon further cooling. This indicates 
structural fluctuations by softening of the lattice or vibrational modes above the structural
distortion which is also mirrored in other properties (e.g. thermal
expansion\cite{Klingeler2010} and \xray\
reflection broadening\cite{McGuire2008}). 
The angles with the angle bisector
parallel $a_\text{o}$ and $b_\text{o}$ do not exhibit this anomaly and merely
differentiate corresponding to the tetragonal to orthorhombic transition.

Although, the overall structural changes upon cooling are
subtle, they indicate a response of the structure already at temperatures above
the onset of the metric distortion. It is clear that higher resolution of the
structural investigations
is needed to further study the interplay of structure and physical properties
of both the ``parent'' and superconducting phases.

\section{Conclusion}
A geometrical model
was developed for single-crystal synchrotron \xray\ diffraction measurements to
quantify the metric distortion and twin formation of GdFeAsO induced by the
tetragonal to
orthorhombic structural transition.

Subtle changes in the structure are detectable. 
Fitting the spontaneous deformation by a power law lead to a small
critical
exponent of $0.072(1)$, which is lower than expected for a two-dimensional
Ising model, pointing towards a transition close to first order. This may be
caused by the separation of the magnetic and the structural transition in
temperature. 
The lattice parameters decrease with decreasing temperature, and the $a$ and $b$ 
axes differentiate at temperatures below the structural transition. 
The deformation is
abrupt and preceeded by a stagnation of the decrease of the cell parameters.
This indicates long-range fluctuations of the lattice or softening of a
vibrational mode at
temperatures above the structural transition, which were also observed by other
methods (e.g.~Ref.~\onlinecite{Hahn2012}). If this feature is linked to the
superconductivity is not yet clear.

Although the presented structural information from single-crystal \xray\
diffraction show the subtle structural changes around the tetragonal
to orthorhombic phase transition, data with higher resolution is needed to
unreavel more details of the interplay of the structure and properties 
of the ``parent'' compounds of iron-based superconductivity. 

\section{Acknowledgement}
Parts of this research were carried out at the light source DORIS III at
DESY, a member of the Helmholtz Association. We would like to thank Martin
Tolkiehn and Mathias Herrmann for their assistance at beamline D3.  The authors
also like to thank Jutta Krug, Karoline Stolze, Eike Ahrens for their help with
sample preparation and measurements.


%

\end{document}